\documentstyle[12pt,aasms4]{article}

\input psfig

\renewcommand{\thebibliography}[1]{\clearpage\subsection*{REFERENCES}\list
 {\arabic{enumi}.}{\settowidth\labelwidth{[#1]}\leftmargin\labelwidth
 \advance\leftmargin\labelsep
 \usecounter{enumi}}
 \def\newblock{\hskip .11em plus .33em minus .07em}
 \sloppy\clubpenalty4000\widowpenalty4000
 \sfcode`\.=1000\relax}

\def\approxlt{\lower.2em\hbox{$\buildrel < \over \sim$}}

\def\ha{\ifmmode {{\rm H}\alpha}
        \else {H$\alpha$}\fi}
\def\hnought{\ifmmode H_0
    \else $H_0$\fi}

\def\la{\ifmmode {{\rm Ly}\alpha}
        \else {Ly$\alpha$}\fi}

\def\p.{^{\prime\prime}\kern-2.1mm .\kern+.6mm}
\def\pone{^{\prime}\kern-1.05mm .\kern+.3mm}
\def\qnought{\ifmmode q_0
    \else $q_0$\fi}
\def\sqr#1#2{{\vcenter{\hrule height .#2pt
        \hbox{\vrule width .#2pt height#1pt \kern#1pt
                \vrule width .#2pt}
        \hrule height.#2pt}}}

\def\ten#1{\ifmmode 10^{#1}
    \else $10^{#1}$\fi}
\newcount\hours
 \newcount\minutes
  \def\SetTime{\hours=\time
         \global\divide\hours by 60
         \minutes=\hours
         \multiply\minutes by 60
         \advance\minutes by-\time
         \global\multiply\minutes by-1 }
 \SetTime
 \def\now{\number\hours:\ifnum\minutes<10 0\fi\number\minutes}

\def\b{\bigskip}
\def\p{\par}

\def\ni{\noindent}

\begin{document}

%\rightline{\bf DRAFT: PLEASE DO NOT QUOTE}

%\leftline{printed \Now}
%\rightline{version 01-jul-1996}

\title{\large\bf Molecular gas and  dust
around a radio-quiet quasar at redshift 4.7}
\author{Alain~Omont \altaffilmark{1}, Patrick~Petitjean \altaffilmark{1,2},
St\'ephane~Guilloteau \altaffilmark{3}}
\author{Richard~G.~McMahon \altaffilmark{4}, P.~M.~Solomon \altaffilmark{5},
Emmanuel~P\'econtal \altaffilmark{6}}

\vskip 1.0truecm

{
\small
\baselineskip 10pt 
$^1$ Institut d'Astrophysique de Paris--CNRS, 98bis Bd Arago,
 F--75014 Paris, France. \hfill\break
$^2$ DAEC, URA CNRS 173, Observatoire de Paris-Meudon,
 F-92195 Meudon, France. \hfill\break
$^3$ Institut de Radio Astronomie Millim\'etrique, 
F--38460 St Martin d'H\`eres, France. \hfill\break
$^4$ Institute of Astronomy, Madingley Road, Cambridge CB3 0HA, 
 UK. \hfill\break
$^5$ Astronomy Program, State University of New York, Stony Brook,
NY 11794--2100, USA. \hfill\break
$^6$ Centre de Recherche Astronomique de Lyon, CNRS UMR 142,
9 Avenue Charles Andr\'e, F--69561, St Genis Laval, France.
}
\p

\vskip 1.0truecm

\begin{center}
email: omont,petitjean@iap.fr, guillote@iram.fr,
rgm@ast.cam,ac.uk, psolomon@sbastk.ess.sunysb.edu,
pecontal@obs.univ-lyon1.fr
\end{center}

\vskip 2.0truecm 
\centerline{To appear in Nature, Aug 1st 1996}

\newpage

{\bf Galaxies are believed to form a large proportion of their stars
  in giant bursts of star formation early in their lives, but when and
  how this took place are still very uncertain.  The presence of very
  large amounts of dust in objects at the largest known redshifts,
  quasars and radiogalaxies at z$\geq$4$^{1-6}$ shows that some
  syhthesis of heavy elements has already occurred at this time.  This
  implies that molecular gas--the building material of stars-- should
  also be present as it is in galaxies at lower redshifts$^{7-10}$.
  Here we report the detection of emission from dust and carbon
  monoxide in the radio-quiet quasar BR1202--0725 at redshift z=4.7.
  Maps of these emissions reveal two objects separated by a few
  arc seconds, which could indicate either the presence of a
  companion to the quasar or gravitational lensing of the quasar
  itself. Regardless of the precise interpretation of the maps,
  the detection of carbon monoxide confirms the presence of a large
  mass of molecular gas in one of the most distant galaxies known,
  and shows that conditions conducive to the huge bursts of star
  formation existed in the very early Universe.
}

%\newpage

We used the 4 15m-antenna interferometer of the Institut de Radio
Astronomie Millim\'etrique (IRAM) on the Plateau de Bure, France, in
Dec. 1995 and Jan. 1996, for a total integration time of $\sim$ 16
hours with projected baselines from 15m to 160m. Simultaneous
measurements were made at $\lambda$ = 1.35 and 2.9mm for the purpose
of measuring the dust continuum and the CO(5-4) line respectively.
The redshift coverage, centered at the approximate redshift of the
Ly$\alpha$ line of the companion$^{12}$, is 4.679 to 4.730 for the
CO(5-4) line (velocity range 2700 km/s). The equivalent single band
system temperatures were $\sim$120--150K and $\sim$300K at 3mm and
1.35mm respectively.  Additional data was obtained in Apr. 1996 at
$\lambda$ = 1.35mm, and at 3.7mm to measure the CO(4-3) line over
redshift range 4.675 to 4.709.

The 1.35mm continuum emission consists in an elongated source at the
position of the quasar, and a second component about 4" to the
North-West with $\sim$35\% of the total flux.  Fig. 1 displays the
various features detected in the visible superposed on the millimetre
map. The coincidence of the visible position Q of the quasar with the
main millimetre peak is almost perfect and well within the
uncertainty, $\sim 0.5''$, of the relative position of the quasar with
respect to the millimetre map frame.  The millimetre emission is
extended in the NW direction where a companion to the quasar has been
detected in the optical and near--infrared broad-band images$^{11-14}$
However, it is seen in Fig. 1 that there is definitely no coincidence
between these companion visible features and the NW 1.35mm peak. \p

Spectra in the redshift range 4.679-4.704 of the CO(5--4) line,
with a velocity resolution of 60
km.s$^{-1}$, are shown in Fig. 2 at various positions. Line emission is
detected towards both sources, with an integrated
intensity of  1.1$\pm$ 0.2 Jy.km.s$^{-1}$ towards the quasar and
$1.3 \pm 0.3$ Jy.km.s$^{-1}$ towards the NW peak.
However, the actual linewidth is uncertain in the latter position.
No signal is detected at other positions in the map (Fig.~1d, 1e and 2).
The displayed fits yield a FWHM=190~km/s and a
central z=4.6947 at Q, and 350~km/s and z=4.6916 at NW.
The derived widths are  in the range observed for other high redshift
sources: 220 km/s for F10214+4724$^9$, 326 km/s in
H1413+117$^{10}$, from 150 to 550 km/s for a sample of 37
ultraluminous galaxies$^{7,15}$.\p

Each of these lines is detected at $\simeq 5-6 \sigma$ level.
Since both detections are independent because the two
positions are spatially resolved, CO(5-4) emission is definitely
detected at about the $8 \sigma$ level. The coincidences with the two
1.35mm peak positions and the absence of signal elsewhere (Fig.~1de) further
strengthen this conclusion. The lack of pure continuum emission when all the
3mm broad band measurements, except in the line frequency range, are included
indicates the detected 3mm emission comes from a spectral line (Fig.~1b).\p

The presence of CO is further confirmed by the detection of other lines. The
3.7~mm J=4-3 line is detected at the 5-6$\sigma$ level 
with integrated flux 1.5$\pm$0.3~Jy~km~s$^{-1}$ (see Fig 1e). 
No continuum was detected at this frequency (Fig 1c). 

Moreover,
the CO(7-6) line (Fig. 2) is present at the 3$\sigma$ level in a spectrum taken
in November 1993 with the IRAM 30m telescope at Pico Veleta Spain. Its
integrated   flux (3.1$\pm$0.86~Jy~km~s$^{-1}$), central redshift (4.6915 $\pm
.001$) and width ($\sim$250--300~km/s) are consistent with the 5--4 line
detected at Bure, taking into account the large 2mm beam of the 30m (17$''$)
and excitation conditions similar to F10214+4724$^{9}$ and H1413+117$^{16}$. The
observed flux ratio = 3.1/2.4 translates into a brightness temperature (or line
luminosity, L') ratio for the (7-6)/(5-4) lines of  0.65 indicating warm gas.
In order to estimate the mass of molecular hydrogen in BR1202$-$0725, several
approximations are required. The high rotational temperature inferred from the
above ratio indicates that the line luminosity L' will not be much larger for
lower J levels.  Thus,  using the (5-4) line luminosity and a conversion factor
M$(H_2)/L'_{CO}$ = 4 (see ref 9), we obtain $M(H_2) = \, 6 \times
10^{10}M_{\odot}$. This is probably an overestimate;   the H$_2$ mass derived
this way for ultraluminous IR galaxies with similar CO luminosities$^{7,15}$ is
too high by about a factor of 3.

Another high redshift object with similar apparent dust and gas properties is
F10214+4724 at z = 2.3, a lensed infrared galaxy.
Table 1 details the comparison of the emission from BR1202$-$0725 and F10214+4724. It is
seen that the observed 3mm integrated line flux and
luminosity
as well as the apparent molecular mass, are
only a factor $\sim$2 smaller in BR1202$-$0725.
The actual
luminosity and H$_2$ mass depend on an eventual magnification (see below). The
relative strengths of the rest frame far infrared/submm and CO emissions, after
applying a K (redshift) correction for the dust spectrum are strikingly
similar in BR1202$-$0725 and F10214+4724$^{9}$, and also in H1413+117
(Cloverleaf)$^{10,17}$. \p

At z=4.7, the $\sim4''$~mm extension
typically corresponds to a projected distance of $\sim$12--30~kpc.
If there is no strong lensing, there are at least two
distinct far--infrared emitters with their own heating
source (AGN and/or starburst). The mass of dust, given by Eq 1 in ref~$^1$,
should be comparable in both sources and $\sim 10^8~{\rm M}_{\odot}$.
These two
nearby relatively massive galaxies may thus have formed simultaneously,
probably in an early stage of the formation of the central core of a cluster,
and they could rapidly merge in a time
comparable to the time scale of the activity of bright quasars ($\sim$10$^8$~yr,
see e.g.~$^{19}$).
The companion observed in the millimetre continuum could be similar to
F10214+4724,
with most of its energy emitted in the far infrared. However, contrary to
F10214+4724, it
is not detected in the visible. The strong tidal interaction
between both objects can boost both the star-formation and AGN activity.
It is worth noting that the companion,  has a CO redshift
z=4.6915, FWHM=250-300~km/s, very close (with a difference in velocity of
~$\sim$~200~km/s) to that of the metallic absorption system at z=4.688$^{20}$
(FWHM~=~260~km/s~$^{20}$).
The absorption system may be due to the halo of the millimetre galaxy 
companion.
 
 The possibility that the double image is due to gravitational lensing
 should also be considered.  Lensing effects are clear on optical
 images of the other high redhsift QSO's with CO emission, H1413+117
 (Cloverleaf, 4 images$^{22}$) and F10214+4724 (gravitational arc system,
 see e.g.$^{21,22}$ and references therein).  Optical amplification
 factors of $\sim$10 or larger are inferred and for F10214+4724 the
 magnification of the CO emission is of the order of 5--10$^{25,}$.
 The Cloverleaf also clearly shows evidence of lensing in the CO
 emission$^{24}$.  There are some indications of the possibility of
 strong lensing on the line of sight to BR1202$-$0725; B. Fort and S.
 D'Odorico (private communication) have measured a strong
 gravitational shear in the field. The line of sight to BR1202$-$0725 itself
 has strong absorption systems at z=1.75, 2.44 and
 4.38~$^{19,25,26}$.  A good explanation of the absence of the
 second bright optical image could be a strong interstellar absorption
 in the deflector galaxy along one light path only. Such cases of
 strong absorption in lens systems are known (see e.g. ref~$^{27}$).
 However, the faint Ly$\alpha$ companion is unlikely to be the result
 of lensing since it has a spectral energy distribution$^{11-14}$ and
 emission line spectrum different from that of the quasar.

In either case, no lensing or modest magnification of the CO emission, these
observations clearly show, for the first time, the presence of a large mass of
molecular gas and dust   in one of the most distant --  and hence youngest --
quasars.   Combined with the other detections of CO in
quasars
at z $\approx$ 2.5, this  seems to
settle completely the question of whether high-redshift quasars are associated
with galaxies.
    BR1202$-$0725 has a CO luminosity  comparable to  the highest observed
in nearby luminous IR galaxies$^{7, 15}$, which show characteristics of  huge
starbursts and AGN's.
Thus, star formation  in a massive metal enriched ISM,
existed very early in the history of the Universe only 0.7 Gyr after the Big
Bang (H$_0$ = 75, q$_0$ = 0.5).  This supports the idea that the emergence of
quasars at very high redshift is connected with the onset of galaxy formation,
possibly with formation of the core of ellipticals (see e.g. ref~$^{32}$),
although nearby quasars also show strong CO emission.

The companion of BR1202$-$0725, if real, may  be
an infrared galaxy  with very high extinction in the optical and ultraviolet.
If the companion is representative of young galaxies, massive star formation at
very high redshift will be visible primarily at millimeter wavelengths.

\vfill\eject

\def\ref{\par\noindent\hangafter=1\hangindent=1cm}

\ref ~1. McMahon, R.G., Omont, A., Bergeron, J., Kreysa, E., Haslam, C.G.T.,
{\it MNRAS}, {\bf 267}, L9-L12 (1994)

\ref ~2. Isaak, K.G., McMahon, R.G., Hills, R.E., Withington,S.,
{\it MNRAS}, {\bf 269}, L28-L32 (1994)

\ref ~3. Dunlop, J.S., Hughes, D.H., Rawlings, S., Eales, S., Ward, M.,
{\it Nature}, {\bf 370}, 347-349 (1994)

\ref ~4. Chini, R., Kr\"ugel, E., {\it Astron. Astrophys.}, {\bf 288}, L33-L36
(1994)

\ref ~5. Ivison, R.J., {\it MNRAS}, {\bf 275}, L33-L36 (1995)

\ref ~6. Omont, A., McMahon, R.G., Cox, P., Kreysa, E., Bergeron, J., Pajot, F.
and Storrie-Lombardi, L.J., {\it Astron. Astrophys.} in press (1996)

\ref ~7. Solomon, P.M., Downes, D., Radford, S.J.E., and Barret, J.W.,
{\it Astrophys. J.} in press (1996)

\ref ~8. Brown, R.L. and Van den Bout, P.A., {\it Astrophys. J.}, {\bf 397},
L19--L22 (1992)

\ref ~9. Solomon, P.M., Downes, D. and Radford, S.J.E.,  {\it Astrophys. J.}
{\bf 398}, L29-L32 (1992)

\ref  10. Barvainis, R., Tacconi, L., Antonucci, R., Alloin, D., Coleman, P.,
{\it Nature}, {\bf 371}, 586-588 (1994)

\ref 11. Hu, E.M., McMahon, R.G. and Egami, E., {\it Astrophys. J. Lett.}
{\bf 459}, L53-55 (1996)

\ref 12. Petitjean, P., P\'econtal, E., Valls-Gabaud, D. and Charlot, S.,
{\it Nature} {\bf 380}, 411-413 (1996)

\ref 13. Djorgovski, S.G., in {\it Science with VLT}, Walsh, J.R. \&
Danziger, I.J. eds, (ESO, Springer, Berlin), 351 (1995)

\ref 14. Fontana, A., Cristiani, S., D'Odorico, S., Giallongo, E. and
Savaglio, S.,  {\it MNRAS}, in press (1996)

\ref 15. Downes, D., Solomon, P.M. and Radford, S.J.E., {\it Astrophys. J.}
{\bf 414}, L13-L16 (1993)

\ref 16. Barvainis, R., in: {\it Cold Gas at High Redshift},
M.~Bremer, H.~Rottgering, P.~van der Werf \&  C.~Carilli eds (Kluwer) (1996)

\ref 17. Barvainis, R., Antonucci, R., Hurt, T., Coleman, P. and
Reuter, H.-P., {\it Astrophys. J.}, {\bf 451}, L9-L12 (1995)

\ref 18. Haehnelt, M.J., Rees, M.J., {\it MNRAS}, {\bf 263}, 168-178 (1993)

\ref 19. Wampler, E.J., Williger, G.M., Baldwin, J.A., Carswell, R.F.,
Hazard, C. and McMahon, R.G., {\it Astron. Astrophys.} in press (1996)

\ref 20. Magain, P. et al., {\it Nature}, {\bf 334}, 325-327 (1988)

\ref 21. Broadhurst, T. and Lehar, L., {\it Astrophys. J.}, {\bf 450},
L41-L44 (1995)

\ref 22. Eisenhardt, P.R., Arnus, L., Hogg, D.W., Soifer, B.T.,
Neugebauer, G. and
Werner, M.W., {\it Astrophys. J.}, {\bf 461}, 72-83 (1996)

\ref 23. Downes, D., Solomon, P.M. and Radford, S.J.E. {\it Astrophys. J.}
{\bf 453}, L65-L68 (1995)

\ref 24. Scoville, N.Z. et al., in~:~{\it Cold Gas at High Redshift}, M.
~Bremer, H.~Rottgering, P.~van der Werf \& C.~Carilli eds (Kluwer) (1996)

\ref 25. Storrie--Lombardi, L.J., McMahon, R.G., Irwin, M.J., Hazard, C.,
 {\it Astrophys. J.} in press (1996)

\ref 26. Lu, L., Sargent, W.L.W., Womble, D.S., and Barlow, T.A.,
{\it Astrophys. J. Lett.}, {\bf 457}, L1-L4 (1996)

\ref 27. Djorgovski, S.G., Meylan, G., Klemola, A., Thompson, D.J., Weir, W.N.,
Swarup, G., Rao, A.P., Subrahmanyan, R., Smette, A., {\it MNRAS},
{\bf 257}, 240-244 (1992)

\ref 28. Franceschini, A. and Gratton, R., {\it MNRAS} submitted (1996)

\ref 29. Downes, D., Radford, S.J.E., Greve, A., Thum, C., Solomon, P.M.
and Wink, J.E., {\it Astrophys. J.}, {\bf 398}, L25-L27 (1992)

\ref 33. Elston, R., Bechtold, J., Hill, G.J. and Jian G. {\it Astrophys. J. 
Lett.}, {\bf 456}, (1996)

\b\b\ni
ACKNOWLEDGEMENTS. This work was carried out in the context of EARA,
a European Association for Research in Astronomy. RGM thanks the Royal
Society for support. We are grateful to the IRAM staff at Bure for
its efficient assistance. We thank J.~Bergeron, S.~Charlot, P.~Cox,
D.~Downes, B.~Fort, Y.~Mellier S.~Radford and P.~Schneider for useful
discussions. \p 

\vfill\eject

%\documentstyle[12pt,twoside]{article}

%\pagestyle{empty}
%\textwidth=16.5truecm
%\topmargin -1.5truecm
%\oddsidemargin -0.5truecm
%\evensidemargin -0.5truecm
%\baselineskip 1truecm
%\begin{document}

\renewcommand{\arraystretch}{2}

Table 1 -- Properties of BR1202--0725 and F10214+4724

\begin{tabular}{l|c|c|c|c|l|l}
& \multicolumn{2}{|c|}{BR1202--0725} & \multicolumn{2}{|c|}{F10214+4724}
& Multiplier & Units \\
\hline
D$_{\rm L}$ & \multicolumn{2}{|c|}{19} & \multicolumn{2}{|c|}{8.8} & $$ h$^{-1}$ & Gpc \\
S$_{1.35 {\rm mm}}$ & \multicolumn{2}{|c|}{16} & \multicolumn{2}{|c|}{7.7$^{\dagger}$} & & mJy \\
        z   & \multicolumn{2}{|c|}{4.69}   & \multicolumn{2}{|c|}{2.29}     &          &   \\
Line          & CO(5--4) & CO(7--6) & CO(3--2) & CO(6--5) &  & \\
$\nu_{\rm obs}$ & 101.3 & 141.2   & 105.2    & 210.5    &  & GHz \\
S$_{\rm CO} \Delta$V & 2.4 & 3.1  &  4.1$^9$ & 9.4$^9$  &  & Jy km s$^{-1}$ \\
L'$_{\rm CO}$ & 1.5  & 1.0 & 2.6 & 1.5  & $$ 10$^{10}$A$^{-1}_G$h$^{-2}$ & K\,km\,s$^{-1}$pc$^2$ \\
L$_{\rm CO}$ & 9 & 16 & 3.5 & 16 & $$ 10$^7$A$^{-1}_G$h$^{-2}$ & L$_\odot$ \\
M(H$_2$) & \multicolumn{2}{|c|}{0.6} & \multicolumn{2}{|c|}{1.0}  & $$ 10$^{11}$A$^{-1}_G$h$^{-2}$ & M$_\odot$ \\
X(CO/1.3) & \multicolumn{2}{|c|}{2.0} & \multicolumn{2}{|c|}{3.2} & $$ 10$^3$
& km\,s$^{-1}$ \\
\hline
\end{tabular}

%\noindent
Symbols used~: h~=~H$_{\rm o}$/100 km s$^{-1}$ Mpc$^{-1}$; A$_{\rm G}$ is the
(eventual) gravitational magnification for millimetre emission; D$_{\rm L}$ is
the luminosity distance (assuming q$_{\rm o}$~=~0.5); $\nu_{\rm obs}$ is the
redshifted frequency; S$_{1.35{\rm mm}}$ and S$_{\rm CO}~\Delta$V are the
measured 1.35~mm flux density and integrated CO flux respectively;
L$_{\rm CO}$ and L'$_{\rm CO}$ are the CO line luminosities (eqs 1 and
3 of ref. 9); M(H$_2$) is the nominal H$_2$ mass using
M(H$_2$)\,$\propto$\,L'$_{\rm CO}$; the ratio M(H$_2$)/L'$_{\rm CO}$ is assumed
4~M$_\odot$(km~s$^{-1}$~pc$^2$)$^{-1}$ as in ref. 9 for F10214; the ratio
X~=~(1+z)$^{1.5}$~S$_{\rm CO3mm}$~$\Delta$V/S$_{1.35{\rm mm}}$
should be approximately invariant for the same galaxy moved at
different z (see eq. 3 of ref. 9 and eq. 1 of ref. 1). It is thus a good
indication of the relative strength of the FIR/submm and CO emission. The
difference between BR1202 and F10214 is hardly significant taking into account
the calibration uncertainties and a small correction for the differences of
frequency and line excitation.
The total flux measured in the (4-3) line, 1.5 $\pm$ 0.3 Jy\,km\,s$^{-1}$, is in
good agreement with the (5-4) line.

$\dagger$ \ S$_{1.25{\rm mm}}$~=~9.6 mJy from ref. 30, scaled to 1.35mm.

\vfill\eject

\noindent

FIGURE CAPTIONS

\noindent
\underbar{Figure 1} -- 1.35mm and 3mm maps.\p
For each band, two frequency setups were used to cover a total bandwidth of
0.9 GHz around the redshifted frequencies of the CO(5-4) (101 GHz) and
CO(11-10) (223 GHz) lines.
Coordinates are in the J2000.0 system.

Flux calibration was performed by using 3C273 and 3C279 which are
monitored against planets by the IRAM staff; the flux densities were
13-18 Jy (3C273) and 13-16 Jy (3C279) at 223 GHz, with an estimated
uncertainty about 10\%.  
Linear polarisation of 3C279 has been accounted for in the calibration. 
The typical phase noise was 10 to 30 degrees at 223 GHz, and twice better at
3mm. For continuum data, the rms noise is 0.50 mJy/beam at 223 GHz, 0.22
mJy/beam at 101 GHz, 0.45 mJy/beam at 84 GHz.\p

\noindent
\underbar{Figure 1a} -- 1.35mm continuum image: obtained using natural
weighting. The angular resolution is $2.0''\times 1.3''$, at PA=12$^0$.
Contour step is 1.0 mJy/beam or 2$\sigma$. \p
\noindent
Superposed are the positions of the different visible features~:~
Cross Q -- QSO $\alpha$(1950) = 12h 02m 49.26s $\pm$ 0.05s~$\delta$(1950)=
$-{\rm 07}^o$ 25' 51.0" $\pm$ 0.5" measured from the APM data. \p
Companion features with respect to the QSO$^{13}$~:~
Filled Triangle : Continuum feature --
Filled Square : Ly$\alpha$ peak --
Open Triangle : Second continuum feature.\p

The total integrated flux density is 16$\pm$ 2~mJy,
and the deconvolved size of the elongated source around Q is
$3.0'' \times 0.9'' \pm 0.7''$ at PA $120\pm10^0$.\p 

Previous continuum measurements  with the IRAM 30~m (full width at half maximum
 FWHM = 11")
 at 240~GHz are 10.5$\pm$1.5~mJy$^1$
and 12.3$\pm$2.3~mJy$^6$.
The measurements are done at different frequencies however. Given the very steep
dust spectrum, the flux measured with the interferometer at 223~GHz (1.35mm) should
correspond to a flux $\sim$ 25~\% smaller than at 240~GHz (1.25mm); but this
should be approximately compensated
by the possible small contribution of the CO(11-10) line. Altogether,
given the calibration uncertainties of single measurements, especially with
the 30m ($\sim$ 25\%), the 30m and Bure determinations are
consistent.\p

\b
\noindent
\underbar{Figure 1b} -- 2.9mm continuum image -- Contour spacings are
0.45 mJy/beam (2~$ \sigma$). Contribution from CO line emission
over velocities between --170 and +360 km/s has been avoided.
The angular resolution is $5.0'' \times 2.5''$ at PA = $20^0$.\p  

\b
\noindent
\underbar{Figure 1c} -- 3.7mm continuum image -- Contour spacings are
0.45 mJy/beam (1~$ \sigma$). Contribution from CO line emission
over velocities between --160 and +280 km/s has been avoided.
The angular resolution is $6.4'' \times 2.7''$ at PA = $164^0$.\p  

\b
\noindent
\underbar{Figure 1d} -- Map of the CO(5--4) line integrated flux within the
velocity range + 7 km/s -- +360~km/s, peaking towards the QSO. Contour
spacings are 2~$ \sigma$ (0.32~Jy/beam.km/s, corresponding to 0.9~mJy/beam
for continuum emission). Same angular resolution has Fig.~1b.\p
\b
\noindent
\underbar{Figure 1e} -- Same as Fig.~1c for the velocity range -- 170 km/s --
+ 185 km/s, peaking towards the NW component. \p

\b
\noindent
\underbar{Figure 1f} -- Map of the CO(4--3) line integrated flux within the 
velocity range --160 km/s -- +280~km/s. Contour
spacings are 2~$ \sigma$ (0.26~Jy/beam.km/s, corresponding to 0.6~mJy/beam
for continuum emission). Same angular resolution has Fig.~1c. The unfavorable
beamshape precludes separation of the two components.\p

\b
\noindent
\underbar{Figure 2} -- Spectra of
the CO(5-4) line emission
superimposed to the 1.35mm continuum
image. For each spectrum, the curve
is the best fit gaussian profile
with no baseline removed. \p
\b
\noindent
\underbar{Insert} -- Spectrum of CO(7--6) line observed with the
IRAM 30m telescope. The velocity and redshift scales are identical
for the CO(5-4) and CO(7-6) spectra. \p

\vfill\eject

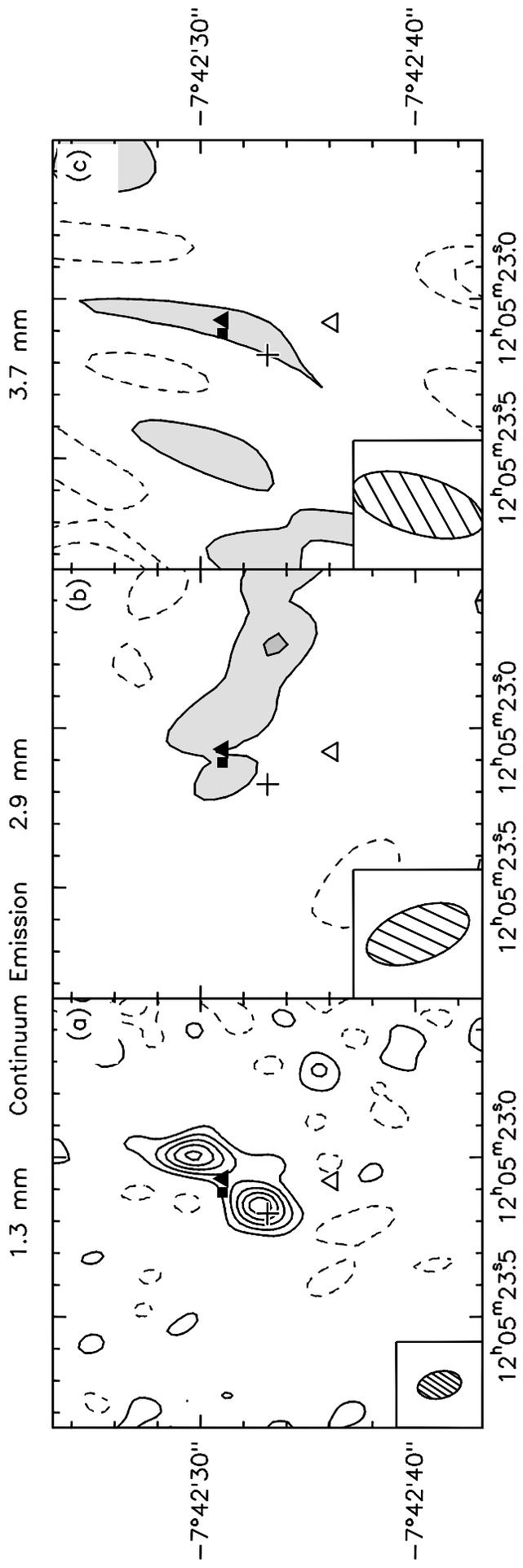
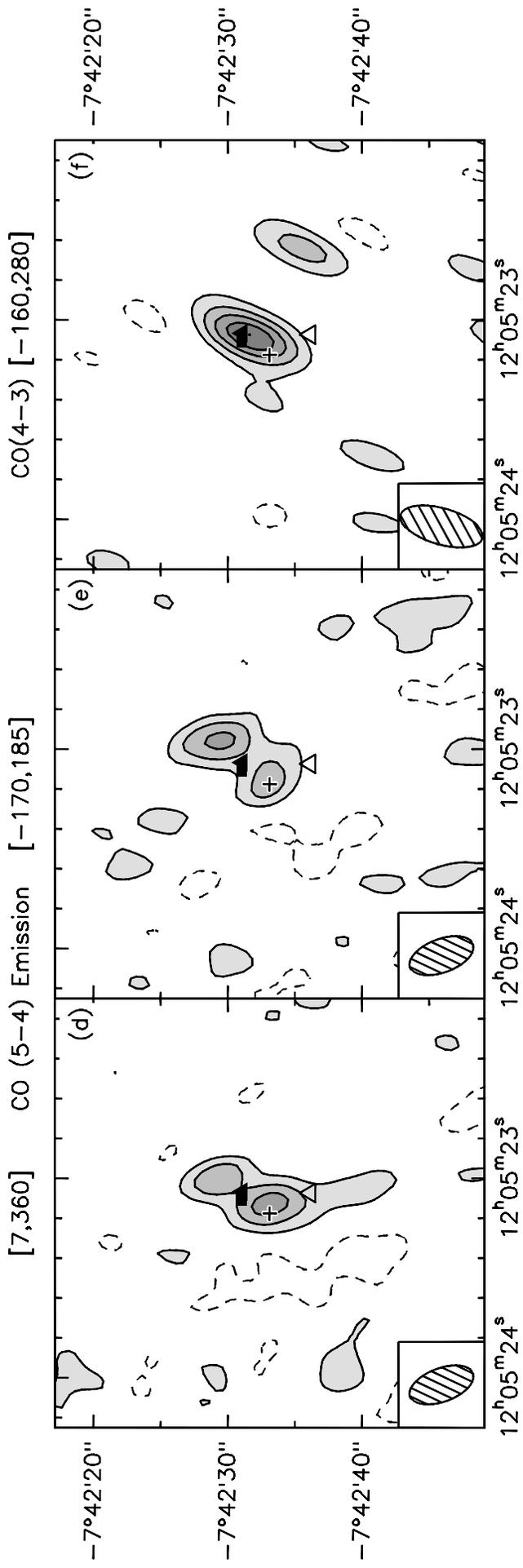
\end{document}